\documentclass{article}[12pt]
\usepackage{authblk}
\usepackage{amsfonts}
\usepackage{amsmath}
\usepackage{amssymb}
\usepackage{bm}
\usepackage{dcolumn}
\usepackage{graphicx}
\usepackage{graphics}
\usepackage[latin1]{inputenc}
\usepackage{latexsym}
\usepackage{rotating}
\usepackage{hyperref}
\usepackage[all]{hypcap} 
\usepackage{xspace} 
\usepackage[usenames]{color}
\usepackage{mathrsfs}
\usepackage[english]{babel}
\usepackage[letterpaper,top=2cm,bottom=2cm,left=3cm,right=3cm,marginparwidth=1.75cm]{geometry}

\newenvironment{acknowledgements}{%
  \begin{abstract}
}{%
  \end{abstract}
}

\usepackage{color}
\usepackage[normalem]{ulem}

\title{A deep classifier of chaos and order in Hamiltonian systems of two degrees of freedom}
\date{}

\begin{document}

\author[1]{Ippocratis D. Saltas \thanks{saltas@fzu.cz}}
\author[2]{Georgios Lukes-Gerakopoulos \thanks{gglukes@gmail.com}}
\affil[1]{CEICO, Institute of Physics, Czech Academy of Sciences Na Slovance 2, 182 21 Praha 8, Czech Republic}
\affil[2]{Astronomical Institute of the Czech Academy of Sciences, Bo\v{c}n\'{i} II 1401, CZ-141 00 Prague, Czech Republic}

\maketitle

\begin{abstract}
Chaos is an intriguing phenomenon that can be found in an immense variate of systems. Its detection and discrimination from its counterpart order poses an interesting challenge. To address it, we present a deep classifier capable of classifying chaos from order in the discretised dynamics of Hamiltonian systems of two degrees of freedom, through the machinery of Poincar\'{e} maps. Our deep network is based predominantly on a convolutional architecture, and generalises with good accuracy on unseen datasets, thanks to the universal features  of a perturbed pendulum learned by the deep network. We discuss in detail the significance and the preparation of our training set, and we showcase how our deep network can be applied to the dynamics of geodesic motion in an axi-symmetric and stationary spacetime of a compact object deviating from the Kerr black hole paradigm. Finally, we discuss current challenges and some promising future directions.
\end{abstract}

\section{Introduction}
 
 Chaos is a fascinating phenomenon which underlines the non-linear nature of dynamical systems, from finance \cite{Kalpinelli:2016e} and biology \cite{Kosmou:2021c} to astrophysics \cite{Contopoulos:2002b} and quantum mechanics \cite{Carlos:2019PhRvL}. It is absent within integrable systems, that is, systems which possess the same number of independent and in involution integrals of motion as their degrees of freedom. However, realistic systems are usually far from being integrable, due to the non-linear interactions present in them. Still, they can be typically described using perturbation theory around integrable systems.

There are theorems describing the transition from a Hamiltonian integrable system to a non-integrable one. The most famous one is probably the {\em Kolmogorov-Arnold-Moser} (KAM) theorem \cite{Kolmogorov54,Arnold63,Moser62}. This theorem implies that under a small perturbation of an integrable system tori, which are sufficiently away from resonances, will survive the perturbation, but they will be deformed. On the other hand, at resonances the tori dissolve. In particular, for a two degrees of freedom system, the Poincar\'e-Birkhoff theorem \cite{Birkhoff13} implies that from the infinite number of periodic orbits on a resonant torus of an integrable system only an even number will survive the perturbation, half of the orbits will be stable (elliptical) and the other half unstable (hyperbolic) creating a structure known as Birkhoff chain on a Poincar\'e surface of section.

The concept of a Poincar\'e surface of section lies in the heart of our work and it was introduced by Poincar\'e in his pioneering work \cite{Poincare93}. This surface of section is transverse to the Hamiltonian flow, which intersects it in regular time intervals, and practically maps the dynamics of a four dimensional phase space  on a two dimensional surface. It is important to realize that in this way, a continuous dynamical system as a Hamiltonian is essentially reduced to a discrete one, i.e. to a map. Hence, in the literature a Poincar\'e surface of section is often called also a Poincar\'e map and the terms are used interchangeably.  On a Poincar\'e map, one should be able to find the whole spectrum of order and chaos, which spans from regular regions dominated by KAM curves, Birkhoff chains, where islands of stability are surrounded by a chaotic layer, to chaotic seas, i.e. regions of the phase space where chaos dominates. However, recognizing the category to which each orbit on a Poincar\'e map belongs can be tricky.

For example, chaotic layers, when they are very weak, they can resemble order. Contrary, tiny islands of stability with very high period might resemble chaotic orbits. Hence, in this work, we are interested in  understanding whether a machine-learning (ML) network is capable of discriminating between order and chaos in Hamiltonian systems. We shall use a proof-of-concept analysis using 2-dimensional symplectic maps, which are equivalent to Poincar\'e maps. We will show that ML is able to do such a discrimination and in fact, it is enough to train our network with orbits with {\it one representative map}. We choose this map to be the so--called {\it Standard Map} \cite{Chirikov:1971gli}, since its dynamics arises from those of a kicked pendulum. The pendulum dynamics is the backbone of any resonance \cite{Morbidelli02}, which is the place where chaos appears, when the integrability of a system is broken. 

Our deep network will be able to perform a classification between ordered and chaotic orbits in 2-dimensional symplectic maps. For our deep network, we choose an architecture aimed at pattern detection based on a combination of convolutional and dense layers. Our goal is {\it not} to achieve the best possible accuracy with such a network, but rather, to build and test a deep network which is able to distinguish between the key features of order and chaos in Hamiltonian systems with satisfactory accuracy. As we will see, of particular importance in this regard will be the construction of the training set, since it will have to capture all of the salient features of ordered and chaotic orbits.
Our goals can be summarised are as follows:
\begin{itemize}
\item To provide a proof-of-concept analysis for a deep-learning network capable to classify order and chaos in Hamiltonian systems, focusing on 2-dimensional maps.

\item To construct a benchmark training set of ordered and chaotic orbits based on a representative map, which we choose to be the Standard Map, and study the generalisation of the trained network to symplectic maps deduced from two degrees of freedom Hamiltonian systems. We will, therefore, be exploiting the {\it universal features} of chaos and order in this context which will allow our trained network to generalises on unseen systems.
\end{itemize}

We will see that our trained network will be able to score on average an accuracy of about $85 \%$ on new, previously unseen trajectories of 2-dimensional symplectic maps. This is an accuracy which can be improved, and we will discuss the challenges in this regard. The concept of a deep order-chaos classifier can have wide and interesting applications in many fields spanning from environmental sciences \footnote{See, for example, the Lorenz chaotic dynamical system serving as a basic model for weather prediction.} to finance \cite{KLIOUTCHNIKOV2017368}. In the context of Hamiltonian systems, it can provide an automatic pipeline to classify the system's dynamics, and in particular, within regimes where an 'by-eye' analysis is tedious and by ignoring an exhausting treatment, we could be led to erroneous results or claims (see, e.g., the discussions in \cite{Lukes:2012PhRvD,Lukes:2016}). Our code can serve also as pedagogical tool, since by using it one can learn to discern chaos from order on a Poincar\'e map, or as a benchmark for similar future attempts.

{\it The rest of the paper is structured as follows:} Section~\ref{sec:Hamiltonian-chaos} describes the way chaos emerges within Hamiltonian systems, we present the maps we will be using, and explain how we construct our training dataset. Section~\ref{sec:Analysis} proceeds with the description of our deep network, its training, and testing. Finally, our results are summarised and discussed in Section~\ref{sec:Summary}. The architecture of our deep network is overviewed in Figure \ref{fig:architecture}. 
\\ \\
The Python code together with the required supporting files to re-produce our results are {\bf available online at}:  https://zenodo.org/records/10593869

\section{Chaos in Hamiltonian systems and symplectic maps} \label{sec:Hamiltonian-chaos}

In an autonomous Hamiltonian system, the Hamiltonian has the form $H = H({\bf x})$, where ${\bf x} \equiv (q_i, p_i)$ the $2d$ position momentum variables. Then, for a phase space function $I=I({\bf x},t)$ it holds that  
\begin{align}
\frac{d I}{dt} = \frac{\partial I}{\partial t} + \{ I, H({\bf x})\},
\end{align}
with $\{ \ldots \}$ denoting the Poisson bracket and $t$ the evolution parameter. Hence, the equations of motion read
\begin{align}
\frac{d {\bf x}}{dt} = \{ {\bf x} , H({\bf x})\}.
\end{align}

If $I$ is a constant of motion, then  $d I/dt = 0$. In order the constant $I$ to be in involution with another constant $K$, then $\{ I,K \}=0$. In a Hamiltonian system each independent and involution constant of motion reduces the number of degrees of freedom by one, which implies that the dimension of the phase space is reduced by two. Hence, when the number of degrees of freedom matches the number of integrals, the system is reduced essentially to a one degree of freedom system.  A one degree of freedom Hamiltonian system is always integrable.

The symplectic property of a Hamiltonian system also implies that the volume of a closed surface is not only preserved in the phase space, but also on sections of the phase space. This means that the surface of a closed area on a Poincar\'e map $P$ is preserved as well. Actually, Hamiltonian systems can be discretised and symplectic mappings can be produced, see, e.g., \cite{Chirikov:1971gli,Meiss:1992RvMP}. The discreet dynamics of the system is then dictated by the map
\begin{equation} \label{eq:map}
{\bf x}_{n+1} = {\bf F}({\bf x}_{n}),
\end{equation}
where ${\bf F}$ is in principle a non-linear function of its arguments, and $n$ an integer denoting the n-th depiction of the initial condition ${\bf x}_0$.

Our work considers only non-integrable Hamiltonian systems of two degrees of freedom, which implies that we consider two dimensional mappings. For a two dimensional mapping there are only two characteristic frequencies, one for each degree of freedom. At a resonance, the ratio of these frequencies is a rational number. The KAM tori correspond to irrational numbers of the frequency ratio and the orbits on them are called quasiperiodic. Quasiperiodic orbits appear as smooth curves on a map and in the case they are nested around a stable periodic point, their shape is practically elliptical. These curves are invariant, since a mapping of such a curve reproduces the curve itself. At resonances lie the Birkhoff chains, around each stable period point of such a chain are nested invariant curves forming the so called islands of stability. These islands of stability are embedded in chaotic layers. These layers are results of non-trivial tangle of asymptotic manifolds originating from the unstable periodic points of the chain. There are no frequencies corresponding to a chaotic layer, hence the Fourier spectrum of a chaotic orbit is continuous. On the other hand, the Fourier spectrum of a regular orbit (periodic or quasiperiodic) is discreet providing its frequencies. In practice, even for a chaotic orbit discreet frequencies might appear superimposed on the continuous background, since the orbit might stick for sufficiently long time around regular orbits mimicking their regular behavior. This phenomenon is known as stickiness \cite{Contopoulos:2002b}. For ML network, weak chaotic layers and stickiness might be very challenging as for people.

\subsection{The training set: The Standard Map} \label{sec:SM}

\begin{figure}[htp!] 
\begin{center}
\includegraphics[width=0.45\linewidth]{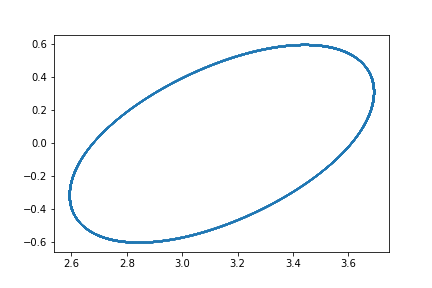}
\includegraphics[width=0.45\linewidth]{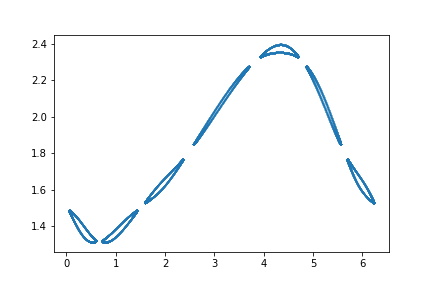} \\
\includegraphics[width=0.45\linewidth]{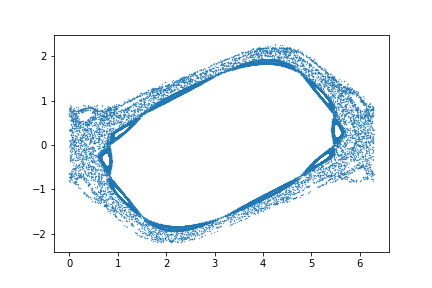} 
\includegraphics[width=0.45\linewidth]{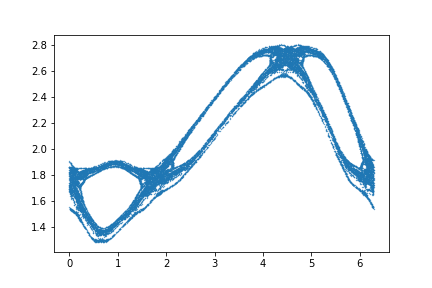}   
\end{center}
\caption{Example orbits for our training set, the Standard Map, defined in equation \eqref{eq:SM}. The Standard Map corresponds to the archetypal 2-dimensional map capturing all the essential features of order and chaos based on a perturbed pendulum. The upper (lower) ones correspond to order (chaos), and the initial conditions correspond to different choices of the constant $K$ and the initial value of $x,y$ (horizontal,vertical axis) on the 2-dimensional plane -- For convenience, we provide their values in the accompanying Python notebook. The upper left plot shows a typical quasi-periodic orbit and the upper right a typical chain of stability islands. The lower plots showcase typical examples of chaos as it emerges around the unstable, asymptotic manifolds in between the chain of stability islands. }
\label{fig:SM}
\end{figure}

The Standard Map is a 2-dimensional discreet map resulting from discretization a Hamiltonian system describing a cyclotron modeled by a kicked pendulum (rotator) \cite{Chirikov:1971gli}. It belongs to the family of twist maps, and it is a key map for both classical and quantum chaos since it describes the emergence of chaos in around the separatrix of non-integrable systems. The Standard Map exhibits all the essential features exhibited by all two dimensional symplectic mappings, and represents in this way the archetypal map of these systems, hence its name. This is the reason we choose it as our training set. The universal features of order and chaos in Hamiltonian systems will allow to use the trained model in other systems/Poincare maps. The Standard Map is defined as 
\begin{align}
& x_{n+1} = x_{n} - K \sin y_n, \nonumber \\
& y_{n+1} = y_n + x_{n+1}, \label{eq:SM}  
\end{align}
with $K$ a real, free constant parametrising the size of the perturbation driving the system away from integrability. The variables $x, y$ relate to angular position and angular momentum and the map has a fixed point at $x = \pi, y = 0$. For $K = 0$, the map corresponds to an integrable system, exhibiting only periodic or quasi-periodic orbits. As the magnitude of $K$ increases, the invariant curves gradually break, homoclinic chaos is turning into heteroclinic chaos, a chaotic sea appears around a main island of stability and after a certain value of $K$, the phase space is dominated by the chaotic sea \cite{Meiss:1992RvMP}. Example trajectories of the Standard Map are shown in Figure~\ref{fig:SM}.
\\

Our {\bf training set is based on the following types of orbits}: 
\begin{itemize}
\item Quasiperiodic orbits (regular orbits)
\item Invariant curves of Birkhoff chain stability islands (regular orbits)
\item Chaotic layers at resonances bounded by invariant curves (chaotic orbits)
\item Chaotic sea (chaotic orbits)
\end{itemize}
We choose to represent equally regular and chaotic orbits. The orbits are produced by iterating $N$-times the Standard Map~\eqref{eq:SM} for a given initial condition $(x_0,y_0)$ and value $K_0$.

The orbits are fed into the network as 2-dimensional images with equal ``pixels" on both axes. As a trade-off between computational efficiency and accuracy, we choose a resolution of $(30 \times 30)$. In order to achieve this, the number of iterations in the computed orbit need to satisfy $30^2 = N_{\text{total}}$, where $N_{\text{total}} = 2 \cdot N$, with $N$ the number of points along each axis respectively ($x,y$). This means that $N = 450$. We produce $1000$ regular orbits and $1000$ chaotic respectively, evenly split amongst the above qualitative sub-cases, by appropriately tuning $(K, x_0, y_0)$. For a given pair of initial conditions $(x_0, y_0)$, we allow $K$ to vary between a range of values for which the map exhibits the desired behaviour (e.g quasiperiodic orbits, etc.).  

It is interesting to highlight that the way we feed the orbits into our network disregards the time evolution of the system ($t_{n} \to t_{n+1}$), since the time-ordered sequence of the points on the plane, $(x_{n},y_{n}) \to (x_{n+1},y_{n+1})$, is not preserved when the resulting array is re-shaped into a grid of $(30 \times 30)$ points. One could indeed preserve the time-ordered sequence by choosing to input the ordered, 1-dimensional arrays of $x$ and $y$ as two distinct channels into the network, and perform a similar analysis\footnote{We are thankful to Ondrej Zelenka for pointing this out to us. }.

\begin{figure}[htp!] 
\begin{center}
    \includegraphics[width=0.6\linewidth]{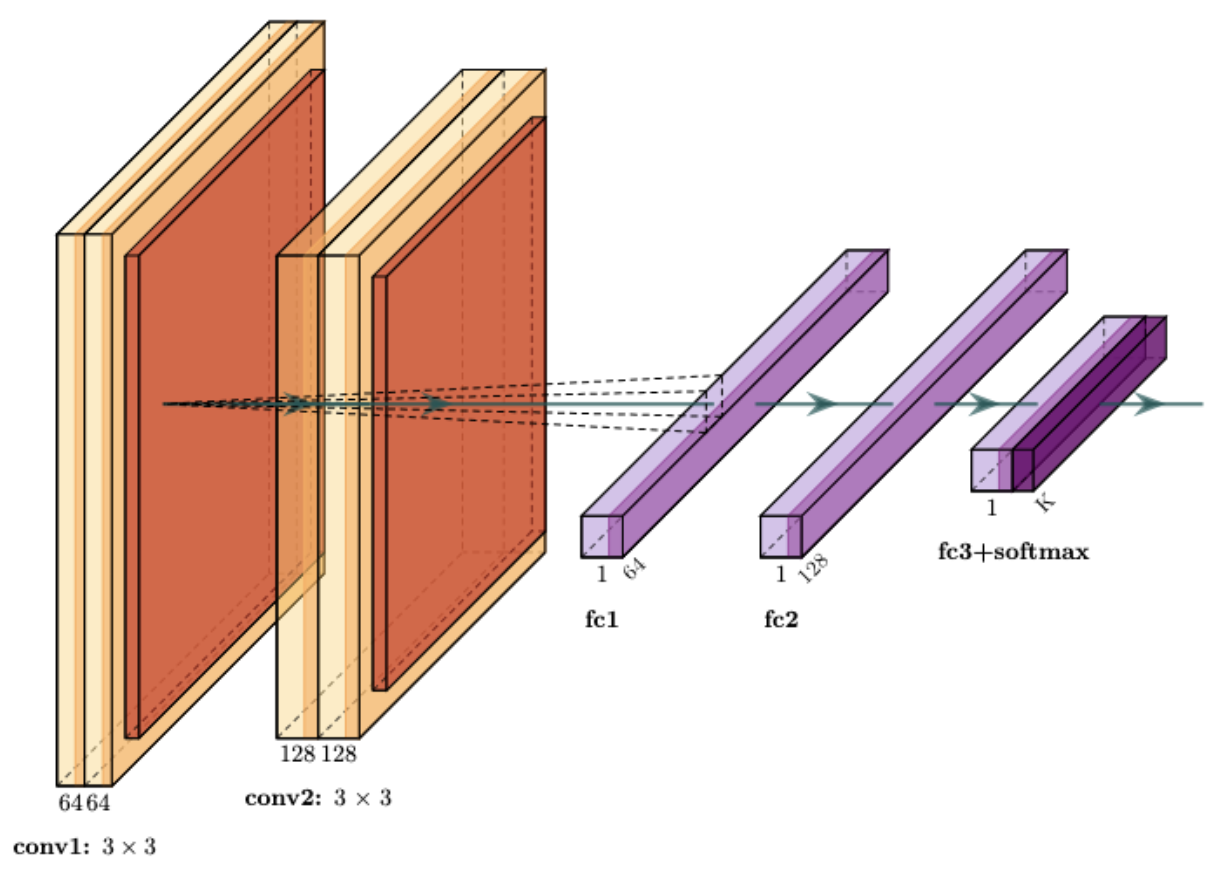} 
\end{center}
\caption{Architecture of our deep classifier. Each stack of convolutional layers (light orange), with $64$ and $128$ filters respectively, consists of $2$ convolutional layers of dimensions $3\times 3$. Each stack is followed by a max-pooling layer with dimensions $2\times2$ (deep orange). The output of the convolutional layers is passed to $2$ dense layers (violet) of $64, 128$ nodes each, with each dense layer followed by a Dropout. All previous layers implement a reLu activation, while the output of the fully-connected layers is batch-normalised and then fed into a soft-max activation function. The output of the network is two real numbers in the range $[0,1]$ corresponding to a probability of the orbit being ordered or chaotic. The training and testing of the model is described in Section \ref{sec:Analysis}.}
\label{fig:architecture}
\end{figure}

\subsection{The testing sets} \label{sec:TestMaps}
We test the performance of our deep network on symplectic mappings and Poincare maps, which the network has not seen before. This provides a test on how well our deep network can generalise to broader systems when detecting order and chaos. We choose to work with the following mappings: the quadratic de Vogeleare map, the Web map, and a map related to geodesic motion in a rotating spacetime. We briefly describe them below, and we refer to \cite{reichl2004} for an overview of the most popular maps related to Hamiltonian systems.

\begin{figure}[htp!] 
\begin{center}
 \includegraphics[width=0.45\linewidth]{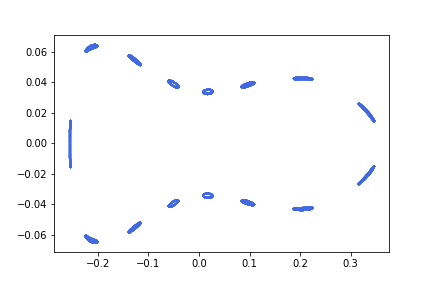}
\includegraphics[width=0.45\linewidth]{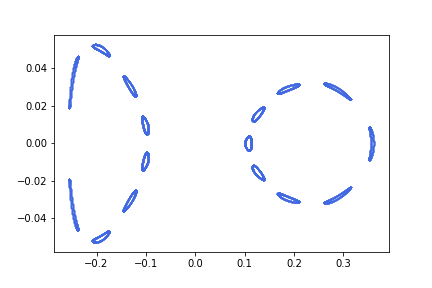}\\
\includegraphics[width=0.45\linewidth]{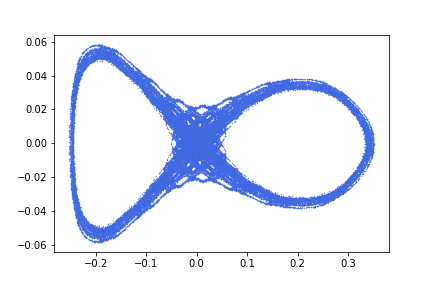}
\includegraphics[width=0.45\linewidth]{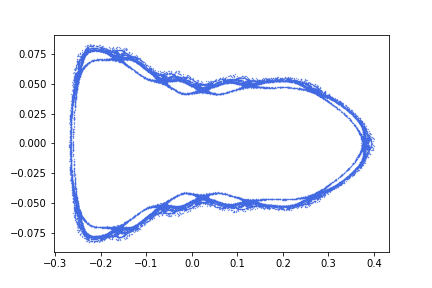}   
\end{center}
\caption{Example orbits for the quadratic deVogelaere map, defined in equation \eqref{eq:deVog}. Our benchmark testing set for the deep classifier is based on this map. We notice that the upper (lower) orbits correspond to order (chaos). Our best-fit deep model is able to perform a $\sim 95 \%$ accuracy on a set of evenly split ordered and chaotic orbits of this map. We remind here that our deep classifier was trained on the Standard Map.   
}
\label{fig:deVog}
\end{figure}

\begin{figure}[htp!] 
\begin{center}
    \includegraphics[width=0.45\linewidth]{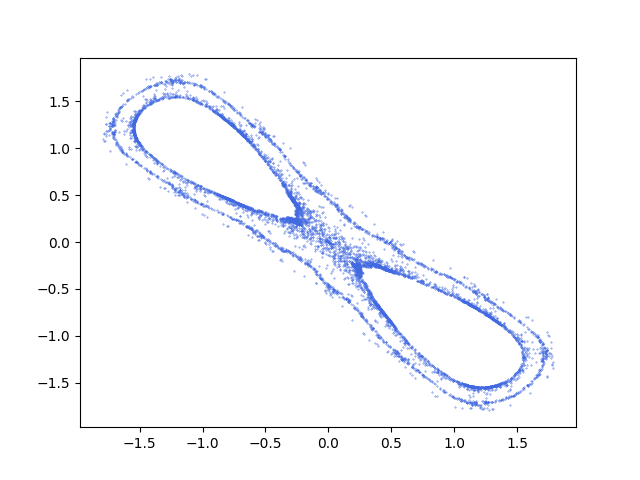} 
    \includegraphics[width=0.45\linewidth]{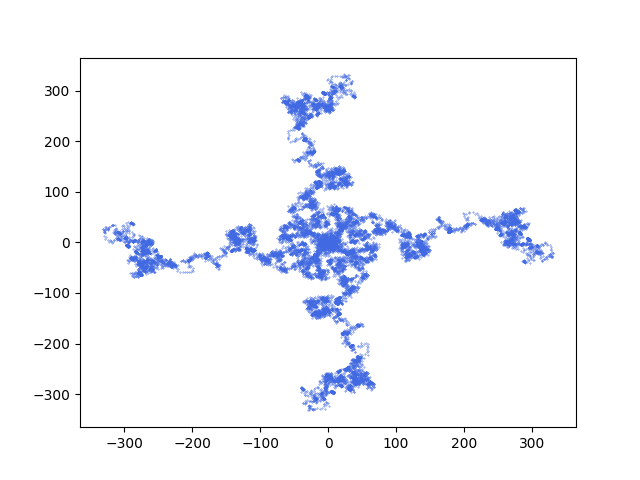} \\
    \includegraphics[width=0.45\linewidth]{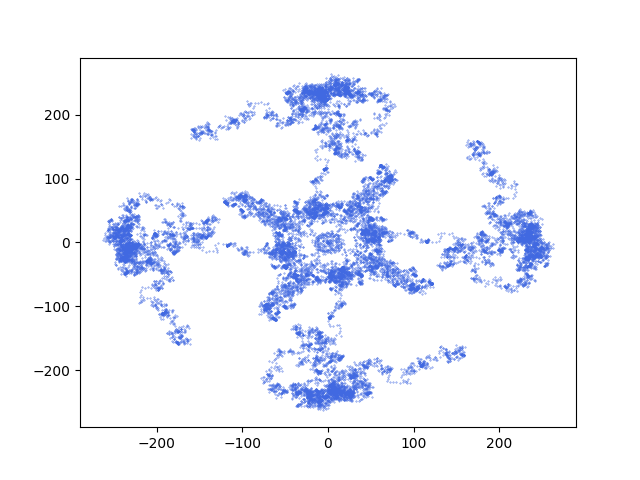}
    \includegraphics[width=0.45\linewidth]{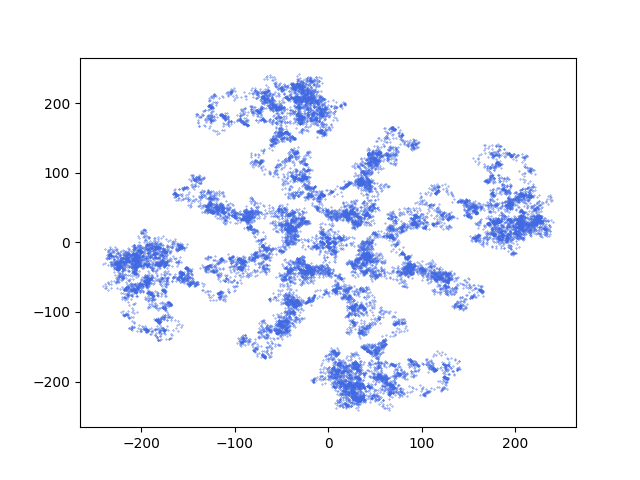}
\end{center}
\caption{Example orbits for the Web map, defined in equation \eqref{eq:Web}, for $q = 4$. As we explain in the text, this map is distinct from the other ones in that the KAM theorem does not apply for this Hamiltonian system. We choose to show only chaotic orbits, due to the particularly interesting web-like structure emerging for this map. This unique structure, which is absent in the other maps studied in this work, results due to the invalidity of the KAM theorem, allowing the formation of the web structure. Each orbit corresponds to different initial conditions ($K, x_0, y_0$), see also the caption of Figure 2. Our best-fit model performs at $\sim 91 \%$ on this map. }
\label{fig:Web}
\end{figure}

\begin{itemize}
\item The {\it quadratic de Vogelaere map} is defined as
\begin{align}
& x_{n+1} = -y_{n} + g(x_n), \nonumber \\
& y_{n+1} = x_n - g(x_{n+1}),  \; \; \text{with} \; \; g(x) = Kx + x^2, 
\label{eq:deVog} 
\end{align}
where, as usual, $K$ is a real perturbation parameter. This map relates to the so--called Quadratic map, and it is equivalent to the quadratic Henon map. Its properties have been explored in detail in \cite{Mackay:1982PhLA} (see also \cite{reichl2004}). Similar to the quadratic map, the quadratic de Vogelaere map is an invertible non-integrable map. It allows for an efficient study of the way chaos emerges in the vicinity of stability islands and its fixed point at $x=0, y=0$ as its stability changes from stable to unstable. Characteristic examples of its phase space are shown in Figure~\ref{fig:deVog}. 

\item The {\it Web map} or Zaslavskii map is a Poincare map defined through \cite{Zaslavskii:1986ZhETF,Zaslavskii:1991wcqr}
\begin{align}
& x_{n+1} =  (x_n + K \sin y_n) \cos\alpha + y_n \sin \alpha, \nonumber \\
& y_{n+1} = -(x_n + K \sin y_n) \sin\alpha + y_n \cos \alpha, \label{eq:Web} 
\end{align}
where $\alpha \equiv 2 \pi/q$, with $q \neq 2$ an integer, and $K$ is a real constant parametrising the perturbation away from integrability. This map arises from a periodically-kicked harmonic oscillator, with the kick exhibiting a particular space dependence. Since the respective Hamiltonian is degenerate, the KAM theorem does not apply.  Different choices of $q$ lead to different resonance structures and symmetries of the phase space.  The morphology of chaotic orbits is drastically different from the other maps, exhibiting a rich set of web-like structures, that resemble fractals. Examples of its phase space are shown in Figure~\ref{fig:Web}. It is easily seen that the map has a fixed point at $x=0, y=0$. 

\item {\it Geodesic motion in strong gravity} 

Here, we are interested in Poincare maps arising from the dynamics of systems in the strong-gravity regime. As our representative system we will use the orbital dynamics arising from geodesic motion within the {\it Johannsen-Psaltis} spacetime, which corresponds to a well-motivated perturbation of the Kerr metric \cite{Johannsen:2011PhRvD}. Contrary to the previous cases, the Poincare maps of this system can be only constructed numerically, since no analytical handle exists \cite{ZLG2017}. Therefore, we will use numerical tabulations of the resulting orbits constructed according to the process explained in Ref. \cite{ZLG2017}. The files with the dataset can be found in the online re-production package. Examples of its phase space are shown in Figure ~\ref{fig:deVog}. 

\end{itemize}

\newpage

%

\begin{figure}[htp] 
\begin{center}
    \includegraphics[width=0.45\linewidth]{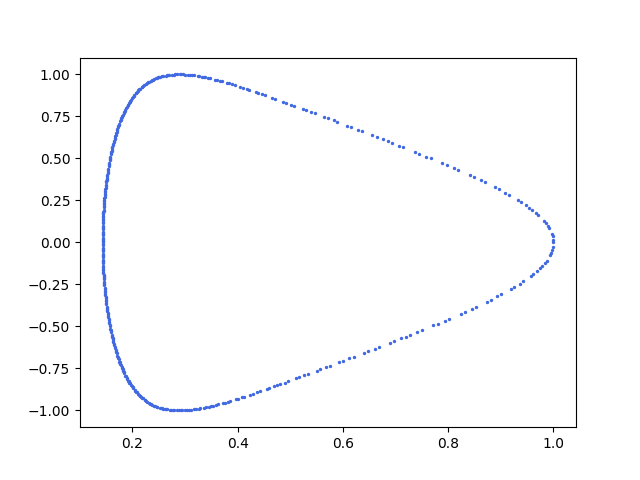} 
    \includegraphics[width=0.45\linewidth]{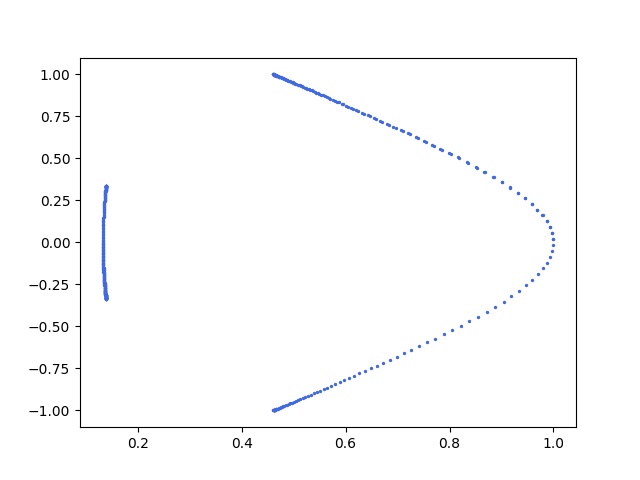} \\
    \includegraphics[width=0.45\linewidth]{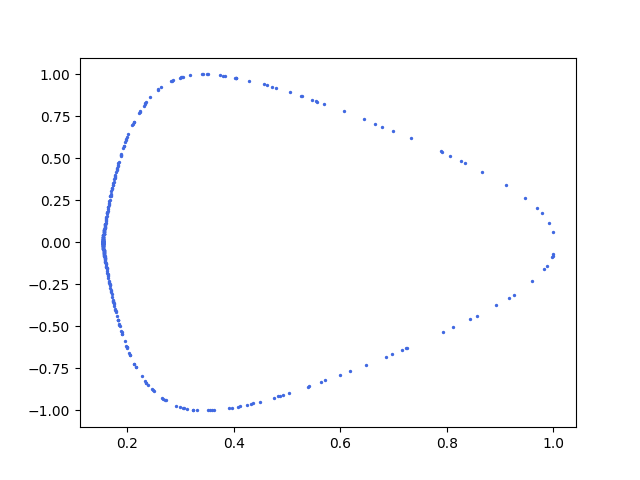}
    \includegraphics[width=0.45\linewidth]{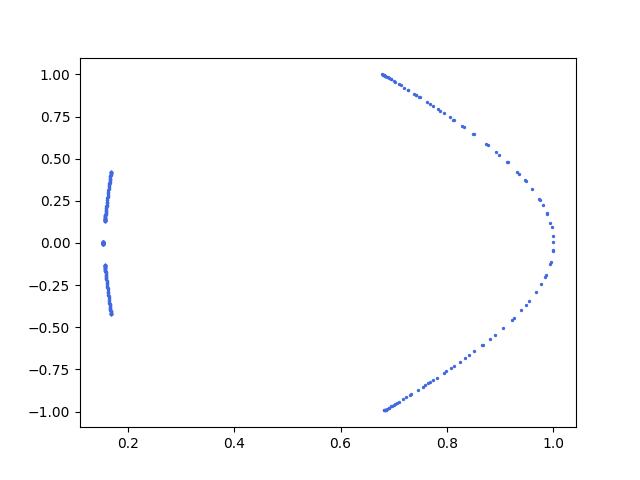}
\end{center}
\caption{Example orbits on Poincare maps arising from geodesic motion in the Johannsen-Psaltis spacetime, as explained in Section~\ref{sec:TestMaps}. As explained in the text, for this case no analytic expressions for the map exist, and the orbits are computed solely by numerical means. The upper (lower) plots correspond to ordered (chaotic) dynamics. Notice the subtle differences between the ordered and the chaotic orbits. Our deep classifier is able to perform a $\sim 90 \%$ accuracy on this data set, which consists in total of $22$ maps. }  
\label{fig:JHMAP}
\end{figure}

\section{Deep network, data augmentation, training and testing} \label{sec:Analysis}

{\bf Architecture and data preparation:} The problem we are trying to solve is a classification one. Given the Poincare map of an orbit, the goal of our deep network will be to classify whether it corresponds to a regular (ordered), or a chaotic orbit respectively. We choose to build a network based predominantly on a {\it Convolutional Neural Network} (CNN) architecture, followed by few dense layers. After experimentation, we found that this architecture is one of the simplest and optimal ones that serves our purpose. The convolutional layers are proven for their ability to detect features, and in our case, they will be equipped with the task of spotting and learning the features associated with order and chaos. Here, we should remind that we are not interested in devising a model with the highest accuracy possible, but rather to present a proof-of-concept of the idea with a good-enough accuracy.  A schematic description of our network is shown in Figure \ref{fig:architecture}. For its implementation we used the {\it Tensorflow} and {\it Keras} libraries - A detailed description of the packages is presented at the end of the paper. 
\\
\\
{\bf Training:} We produced a set of $\sim 2000$ orbits evenly split between order and chaos, based on the Standard Map and according to the recipe described in Section \ref{sec:SM}. We used $2/3$ of this set for training, and $1/3$ for validation. As previously explained, the appropriate representation of ordered and chaotic orbits in the training set is important. We further found that the augmentation of our training set is equally important for the generalisation of our model to new data. We augmented the produced orbits by applying the following operations: we added a small amount of Gaussian noise on each of the computed points in the orbit, and we further applied rotation, shift and shear to the resulting ``image".  We also applied a feature-wise standardisation of the training dataset, which corresponds to an appropriate re-scaling of each image according to the mean ($\mu_{\text{augm.}}$) and standard deviation ($\sigma_{\text{augm.}}$) of the whole dataset. These statistics are computed over all phase spaces of the training dataset, after each one is re-shaped into an $30 \times 30$ image, as explained in Section \ref{sec:SM}. Given the random character of the augmentation process, as a consistency check, we produced $10$ different augmented realisations (the added Gaussian noise on the phase space was kept constant) and found that the scatter of $\mu_{\text{augm.}}, \sigma_{\text{augm.}}$ are small, and in particular, 
$
<\mu_{\text{augm.}}> =  -0.03 \; \;\; (\sigma = 0.04),   \; \; \; <\sigma_{\text{augm.}}> = 0.9 \; \; \; (\sigma = 0.029).
$
This provides a re-assurance that the augmentation procedure yields consistent results. It would be worrisome if the iterative application of augmentation would produce drastically different datasets.
\\
\\
{\bf Testing:} We move on with testing our trained ML network to other area-preserving Poincare maps associated with Hamiltonian systems. As explained in Section \ref{sec:TestMaps} the quadratic deVogelaere map served as our {\it benchmark testing set}. We produced a set of $800$ orbits evenly split between order and chaos, paying attention to retain a similar diversity of orbits as for the training set, described in Section \ref{sec:SM}. The same feature-wise standardisation procedure as with the training set is applied to the testing sets, though the testing sets are not augmented (i.e no application of shear, rotation, etc.). 

Given the inherit stochasticisty in the weights of the trained model (through the random weight initialisation process), as well as in the random character of the augmentation process (e.g random rotation), we are first interested in deriving a mean accuracy and its associated variance. Therefore, we run $10$ times the following iteration: {\bf (i)} produce a new training data set by varying only the random augmentation parameters (e.g rotation, shear) $\to$ {\bf (ii)} train the deep network at fixed hyper-parameters (e.g number of epochs) $\to$ {\bf (iii)} evaluate the trained model on the benchmark testing set. We found that the mean and standard deviation of the test accuracy for our {\it benchmark testing set} (quadratic deVogelaere map) is
\begin{equation}
<\text{test accuracy}> \; \simeq 85\%,  \; \; \; \sigma_{\text{test accuracy}} \simeq 0.06.
\end{equation}
The scatter is low which is re-assuring, while the resulting mean testing accuracy is encouraging and sufficient for our goals. In this sense, the model appears to generalise well on unseen phase spaces. In the produced sample of models according to the above iteration, there were, of course, models scoring higher testing accuracy than the mean. From this set, we selected a model with testing accuracy of $~ 95\%$, as our {\it ``best-fit" model}. 

We would finally like to test the model further, using other two cases described in Section \ref{sec:TestMaps}. The first of these is the so--called {\it Web map}, and the second is the Poincar\'{e} map arising from geodesic motion in the {\it Johannsen-Psaltis spacetime}. For the Web map, we produce a set of $200$ orbits, evenly split between order and chaos, and including chaotic orbits forming the characteristic web pattern. Our best-fit classifier model yields an accuracy of $\sim 91\%$ (see Figure \ref{fig:Web} for example orbits.) 

For the Johannsen-Psaltis  Poincar\'{e} maps (``JHMAP") we use  $22$ orbits, four of these maps are shown in Figure~\ref{fig:JHMAP}. We find that our best-fit classifier model yields an accuracy of $\sim 90\%$. Beyond doubt, this is an encouraging result, taking into account the subtle differences between order and chaos in this system, as shown in Figure \ref{fig:JHMAP}. In particular, the bottom right case in Figure \ref{fig:JHMAP} is very tricky one. Even an expert in the field might characterize it as a regular orbit belonging to a resonance. However, this orbit is chaotic and its evolution takes place inside the resonant islands. 

We close this section with few remarks. We have tested our classification by applying a standardisation procedure based on the given test set. One might wonder what would happen if one had standardised based on the training set. We found that results are still comparable though numerical differences in the results may occur. It is also interesting to comment on the cases where the classifier fails. One issue leading to misleading results is caused by chaotic orbits which are not sufficiently developed due to stickiness, that is, points still remaining confined nearby ordered invariant curves. An example of such orbit is an emerging chaotic orbit around a hyperbolic confined between two stable islands and a KAM curve, after a bifurcation (see, e.g. the top left panel in Figure~\ref{fig:Web}). What is more, the sensitive dependence of chaotic orbits on initial conditions could sometimes be confused with numerical noise. Another issue is the characteristic presence of dense sets of points in chaotic layers again due to stickiness. In regular orbits, possibly due to numerical noise or some geometrical features of the orbit, points may occasionally resemble such a dense set, leading the deep classifier to erroneous results. We leave these issues for future research, and comment on them again in Section~\ref{sec:Summary}.  

\section{Summary and outlook} \label{sec:Summary}

The emergence of chaos in dynamical systems is common in most of the physical systems in Nature, from the dynamics of the solar system \cite{Morbidelli02} to gravitational waves \cite{Lukes:2021hgwa,Destounis:2021PhRvL}. In this work we presented a proof-of-concept pipeline based on deep learning, for the detection of ordered and chaotic dynamics in Hamiltonian systems. Focusing on 2-dimensional Poincare maps, we exploited the universality features of Hamiltonian chaos to train a deep network with the archetypal map, the so--called Standard Map (see Section~\ref{sec:SM}). Our deep network was based predominantly on convolutional layers connected with a set of dense layers (see Figure~\ref{fig:architecture}). 

Key in our analysis has been the creation of a representative training dataset based on the Standard Map, combined with appropriate augmentation, as explain in Section~\ref{sec:Analysis}. We found that our proof-of-concept network generalised well on new, previously unseen datasets, achieving a mean accuracy of $\sim 85 \%$ on our benchmark dataset. Our test sets were based on the quadratic deVogelaere map (benchmark testing set), Web map, and Poincar\'{e} maps arising from geodesic motion in the strong-gravity regime, all described in Section~\ref{sec:TestMaps} (see also Figures~ \ref{fig:deVog}, \ref{fig:Web} and \ref{fig:JHMAP}). 

Our approach can have interesting applications in the future -- in particular, one can use it as a stepstone to discern the different types of regular orbits (e.g., KAM curves vs. islands of stability), to estimate the maximum Lyapunov number \cite{Contopoulos:2002b} for a chaotic orbit or simply provide a more accurate ML scheme discerning chaotic and regular orbits. Our work admits challenging and interesting extensions, which we discuss briefly below:  
\begin{itemize}
\item Chaotic orbits may cover different scales. In particular, they may emerge as as a subtle chaotic layer, within a larger, ordered structure. The detection of such small-scale chaos within an apparent ordered structure is challenging for a deep  classifier. It could be potentially achieved with an adequate combination of convolutions of different dimensions, which could uncover such small-scale features of an orbit on top of the large-scale ones. 
\item Chaos is characterised by an extreme sensitivity on initial conditions. This implies that the numerical integration of such orbits may become very sensitive to small numerical noise, which can in turn lead to drastically different orbits. At the same time, it may happen that a nearby chaotic layer contaminates with numerical noise an apparently ordered orbit, leading to a miss-classification result by the deep network. Therefore, getting a handle upon numerical noise in orbital dynamics is important for a more robust reliability of the deep classifier. 
\item Chaos in Hamiltonian systems exhibits certain universal features, and it was this aspect which we exploited when training our deep classifier. It would be very interesting to explore this universality in the context of deep (conditional) generative networks. For example, through the construction of a deep network that could function as a ``universal" generator of Hamiltonian chaos under certain conditions/requirements. At training, generative deep networks learn the underlying probability density function which degenerates the training data, and are able thereafter to produce different realisations of the data set. Such a deep network could be exploited to study amongst other things the statistical nature of the inherit stochasticity in chaotic orbits. Algorithms such as {\it Generative Adversarial Networks} \cite{goodfellow2014generative} could provide a numerical framework for the realisation of this idea. 
\end{itemize}

\begin{acknowledgements}
IDS is supported by the Czech Academy of Sciences under the grant number LQ100102101, while GLG has been supported by the fellowship Lumina Quaeruntur No. LQ100032102 of the Czech Academy of Sciences. We are grateful to Ond\v{r}ej Zelenka for discussions, feedback, and for providing an independent check of part of our analysis. Computations were done using the following Python libraries: {\tt Numpy 1.24.1, Tensorflow 2.15.0, Keras 2.15.0}. Our code was produced and tested on a {\tt Jupyter} notebook using a machine with a GPU processor. We also acknowledge the use of the {\it Phoebe} computer cluster owned by CEICO/FZU, and we thank its administrator Josef Dvo\v{r}\'{a}\v{c}ek for his valuable assistance. Our code is publicly available at: https://zenodo.org/records/10593869.
\end{acknowledgements}

\bibliographystyle{unsrt}
\bibliography{ref}

\end{document}